\documentclass[conference]{IEEEtran}

\usepackage{cite}
\usepackage{amsmath,amssymb,amsfonts}
\usepackage[linesnumbered,vlined,boxed,commentsnumbered]{algorithm2e}
\SetKwInput{KwData}{Input}
\usepackage{graphicx}
\usepackage{textcomp}
\usepackage{xcolor}
\usepackage{url}
\usepackage{booktabs}

\newcommand{\myref}[1]{\ref{#1}}
\newcommand{\Qtarget}{Q^{\mathrm{target}}}
\newcommand{\Qdiff}{Q^{\mathrm{diff}}}
\newcommand{\Qreduceddiff}{Q^{\mathrm{reduced\,diff}}}
\newcommand{\Qtargetlimitedwalk}{Q^{\mathrm{target}}_{\mathrm{limited\,walk}}}
\newcommand{\Dreduceddiff}{D^{\mathrm{reduced\,diff}}}
\newcommand{\Dtarget}{D^{\mathrm{target}}}
\newcommand{\Ddiff}{D^{\mathrm{diff}}}
\newcommand{\Dwalk}{D^{\mathrm{walk}}}
\newcommand{\dmax}{d_{\mathrm{max}}}
\newcommand{\dmin}{d_{\mathrm{min}}}
\newcommand{\dmaxwalk}{d_{\mathrm{max\,walk}}}
\newcommand{\dfactor}{d_{\mathrm{factor}}}
\newcommand{\qfactor}{q_{\mathrm{factor}}}
\newcommand{\qdifffactor}{q_{\mathrm{diff\,factor}}}
\newcommand{\cglob}{c^{\mathrm{glob}}}
\newcommand{\cpair}{c^{\mathrm{pair}}}
\newcommand{\gmid}{g_{\mathrm{mid}}}
\newcommand{\mumid}{\mu_{\mathrm{mid}}}

\begin{document}
\title{\vspace*{1cm} Interaction-Aware Embedding Optimization for Neutral-Atom Quantum Processors\\
}

\author{\IEEEauthorblockN{Clément de Terrasson de Montleau*}
\IEEEauthorblockA{\textit{Quantum Graph Optimization} \\
\textit{Pasqal}\\
Palaiseau, France \\
clement.de-terrasson@pasqal.com \\
(*corresponding author)}
\and
\IEEEauthorblockN{Victor Drouin-Touchette}
\IEEEauthorblockA{\textit{Faculté de Génie} \\
\textit{Université de Sherbrooke}\\
Sherbrooke, Canada \\
victor.drouin-touchette@usherbrooke.ca}
\and
\IEEEauthorblockN{Wesley Da Silva Coelho}
\IEEEauthorblockA{\textit{Quantum Graph Optimization} \\
\textit{Pasqal}\\
Palaiseau, France \\
wesley.coelho@pasqal.com}
}

\maketitle

\begin{abstract}
     Neutral atom-based quantum computers have the ability to dynamically reshape the qubit register geometry between experimental runs. This capability enables different types of problems to be mapped directly onto the hardware, significantly enhancing the efficiency of solving, for instance, complex optimization and machine learning use-cases. However, this flexibility requires finding atom positions to implement target pairwise interactions, inducing a difficult nonconvex optimization for setting up the quantum machine. In this work, we introduce a new force-directed method that targets interactions to encode the input problem on the quantum computer. Numerical results show that our new method can improve embedding feasibility and  solution quality by orders of magnitude, and remains reliable on the largest instances.
\end{abstract}


\section{Introduction}
\label{sec:introduction}

Optical tweezer arrays of neutral atoms have demonstrated strong potential for analog quantum simulation and the realization of complex quantum Hamiltonians. One of their important features is the ability to engineer arbitrary qubit configurations in both 2D and 3D, together with strong, controllable interactions that can be programmed to suit a wide range of applications \cite{dalyac2024graph}. Once set up, the atom register gives pairwise Rydberg \textit{interactions} that scale as $Q_{ij}\propto D_{ij}^{-6}$ with inter-atom distances $D_{ij}$ in an Ising regime. By carefully adjusting these distances, the interaction strengths can be precisely controlled, enabling the implementation of an interaction Hamiltonian whose structure encodes, for example, the adjacency matrix of a graph for machine learning tasks \cite{djellabi2026kernel}, or off-diagonal terms of a symmetric matrix to solve a quadratic unconstrained binary optimization (QUBO) problem \cite{naghmouchi2024milp}.

The underlying  \textit{register embedding problem} can then be defined as follows: given a target interaction matrix $\Qtarget$, find atom positions so that $Q \approx \Qtarget$ under hardware constraints. Using the notations of Eq.~\ref{eq:qdiff_ddiff}, a natural objective is to minimize a matrix distance $\operatorname{d}(\Qtarget,Q)=\lVert \Qdiff\rVert$. The distance can, for example, be mapped to $\lVert \cdot \rVert_1$ or $\lVert \cdot \rVert_2$. The mapping from geometry to interactions is highly nonlinear, resulting in a non-convex and high-dimensional cost landscape, and making it a difficult problem to optimize.
\begin{equation}
\text{(a)}~\Ddiff = \Dtarget - D \qquad \text{(b)}~\Qdiff = \Qtarget - Q
\label{eq:qdiff_ddiff}
\end{equation}

Existing approaches face some important limitations: distance-based methods do not generally align with the minimization of $\lVert \Qdiff \rVert$ due to the non-linear $d \mapsto d^{-6}$ transformation, while directly applying a black-box optimization can target interactions but is generic by definition and can be sensitive to local minima. For instance, the force-directed Fruchterman-Reingold (FR) algorithm~\cite{fruchterman-reingold} iteratively updates positions to equilibrium by targeting pairwise distances, which is implemented as spring layout in the library NetworkX \cite{networkx-spring-layout} with tunable parameters. The embedding problem can be adapted by converting target interactions $\Qtarget$ into target distances $\Dtarget$, then finding positions whose pairwise distances $D$ minimize $\lVert \Ddiff\rVert_2$. However, this is not generally equivalent to minimizing $\lVert \Qdiff\rVert_2$ (see section \myref{subsec:optimize-interactions}). Additionally, while FR can get trapped in local optima, the situation can be even worse, because the forces do not equilibrate in an optimum of $\lVert \Ddiff \rVert_2$ (see section \ref{subsec:optimize-local-equilibrium}). scikit-learn's Multidimensional scaling (MDS) algorithm~\cite{scikit-learn} can reduce the risk of poor local minima via dimension reduction, and converges to a local optimum using a cost function that can only decrease, but also minimizes $\lVert \Ddiff\rVert_2$ and does not support infinite target distances induced by zeros in $\Qtarget$. SciPy's Nelder-Mead and Broyden-Fletcher-Goldfarb-Shanno (L-BFGS) methods~\cite{scipy} can also be used to solve the register embedding problem by minimizing $\lVert \Qdiff\rVert_2$, but they can remain blocked in local optima. Regarding machine learning related methods, Graph Embedding Autoencoder Network (GEAN)~\cite{GEAN} uses neural networks to embed unweighted graphs, with a Unit-Disk (UD) objective: getting connected pairs closer than non-connected pairs. Finally, \cite{naghmouchi2024milp} assigns atoms greedily on predefined lattice sites, where the sequential choices and the discrete nature of the lattice can make the solutions suboptimal.

In this work we introduce BLaDE (Balanced Latently Dimensional Embedder), a new register-embedding algorithm tailored to encode target matrices (e.g., adjacency matrices and QUBO ones) for neutral-atom based quantum computers. BLaDE schedules exploratory iterations before converging via balanced force weights, uses latent dimensions to escape local minima before returning to physical 2D/3D space, and enforces hardware constraints with a scaling-in-the-loop mechanism designed to preserve solution quality. We benchmark it on representative datasets against those baselines, using interaction-level metrics and constraint feasibility criteria. BLaDE is open-source and available within the QoolQit library~\cite{pasqal_qoolqit}.

\section{Method}
\label{sec:method}

\begin{figure*}
    \centering
    \includegraphics[width=0.75\linewidth]{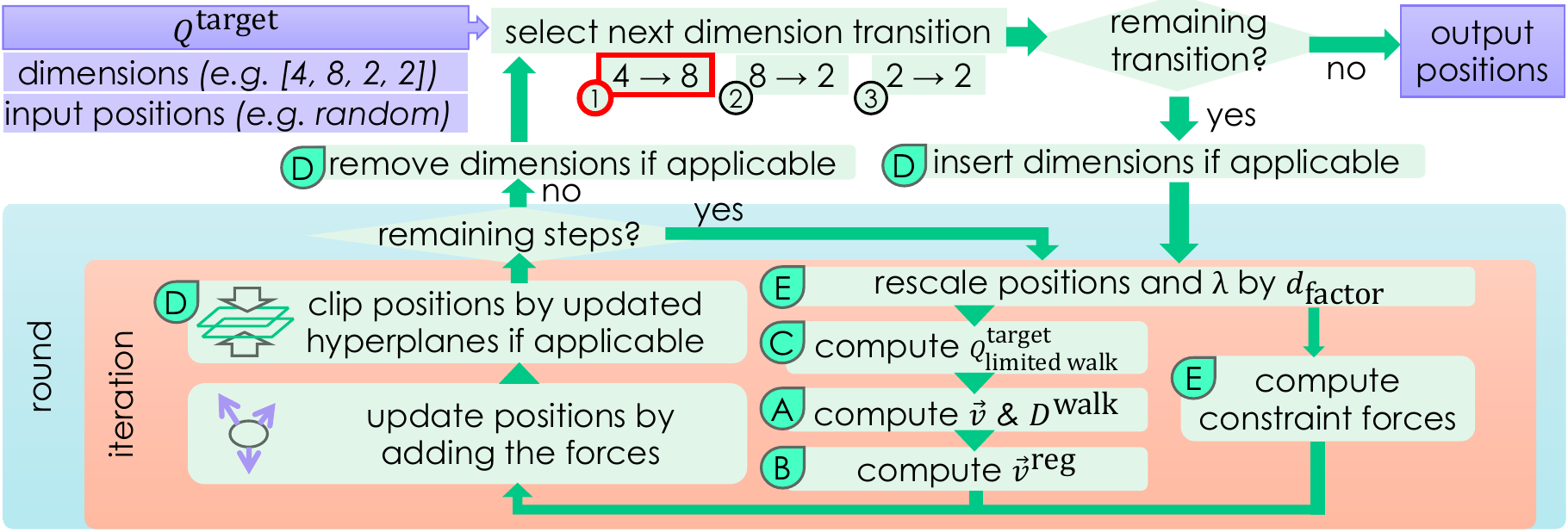}
    \caption{General workflow of BLaDE. Letter tags are described in corresponding subsections of Sec.~\ref{sec:method}. An example of a sequence of dimensions to cross is given starting on the red rectangle transition.}
    \label{fig:workflow}
\end{figure*}

The dynamics of a neutral atom quantum system follows a time dependent Hamiltonian (see \cite{dalyac2024graph}):
\begin{equation}
H(t) = \sum_i \frac{\hbar}{2} \Omega_i(t)\sigma_i^x -\sum_i \hbar \delta_i(t) n_i + \sum_{i<j} Q_{ij} n_i n_j
\label{eq:hamiltonian}
\end{equation}
where $n_i$ is the Rydberg occupation operator of the atom $i$, $\Omega$ and $\delta$ are the amplitude and detuning of the laser, and $\sigma_i^x$ is the Pauli matrix. While the first two parts are controlled using time-dependent laser pulses, the inter-atomic interactions $Q_{ij}= C_6 / D_{ij}^6$ depend on the pairwise distances $D_{ij}$, where $C_6$ is a constant given by the quantum device. Since the operation of the Hamiltonian is given by $\exp\!\left(-\frac{i}\hbar\int_{0}^{T} H(t) dt\right)$, it is invariant under rescaling: applying $H \mapsto \alpha H$ and $t \mapsto t/\alpha$ maintains the same dimensionless dynamics. Consequently, $Q$ and the positions can be rescaled, we set \(C_6 = 1\) without loss of generality, and we work on logical positions of \textit{points}, before rescaling them into positions of atoms.

As in FR, BLaDE takes initial positions (e.g., generated randomly) and computes, at each iteration, ``forces'' that are used as offsets to be applied on the positions. Hardware constraints impose a minimum distance \(d_{\min}\) between any pair of atoms and a maximum radial distance \(d_{\max}\) around the center $(0,0)$. Leveraging the rescaling invariance, these hardware constraints can be reduced to the following logical \textit{distance-ratio constraint}, where $\vec{x}_i$ is the position of point $i$:
\begin{equation}
        \frac{\max_i ~\|\vec{x}_i\|}{\min_{i\ne j} ~D_{ij}} \le \frac{\dmax}{\dmin}
    \label{eq:max_min_distance_constraint}
\end{equation}

In what follows, we give a detailed description of each element of BLaDE, and the workflow is illustrated in Fig.~\ref{fig:workflow}.

\subsection{Optimizing the interactions}
\label{subsec:optimize-interactions}

As in the FR algorithm, the force $\vec{v}_{ij}$ for a pair $(i,j)$ of coordinates $\vec{x}_i, \vec{x}_j \in \mathbb{R}^k$ is directed along $\vec{u}_{ij}=(\vec{x}_j-\vec{x}_i)/\lVert \vec{x}_j-\vec{x}_i\rVert_2$. As depicted in Alg. \ref{alg:applicable_forces}, BLaDE, however, balances the forces' norms for the algorithm to minimize $\lVert \Qdiff\rVert$ instead of $\lVert \Ddiff \rVert$ (see Eq.~\ref{eq:qdiff_ddiff}).  Each force's norm reflects the cost decrease it could make in isolation: to minimize the $L_1$ cost $\lVert \Qdiff\rVert_1$, it is  proportional to $\lvert \Qdiff_{ij}\rvert$, and to minimize the $L_2$ cost $\lVert \Qdiff\rVert_2$ (equivalent to minimizing $\lVert \Qdiff\rVert_2^2$), it is proportional to $\lvert \Qdiff_{ij}\rvert^2$. Here, we consider $L_2$ because it encourages smaller spread errors over concentrated ones. For smoother convergence, we preliminarily reduce the targets through $\qdifffactor \in [0,1]$, which yields the surrogate targets ${\Qtarget}'$ and ${\Dtarget}'$. We also compute $\Dwalk$ as the maximum distance update that can be applied on each pair without overcoming its target.

\begin{algorithm}
\KwData{$\Qtarget$, $D$, $Q$, $(\vec{u}_{ij})_{ij}$\qquad \textbf{Output:} $\Dwalk$; $\vec{v}$}
\BlankLine
$\Qdiff \leftarrow \Qtarget - Q$\;
$\Qreduceddiff \leftarrow \qdifffactor\cdot\Qdiff$\;
${\Qtarget}' \leftarrow Q + \Qreduceddiff$\;
${\Dtarget}' \leftarrow \operatorname{distances}({\Qtarget}')$\;
$\Dwalk \leftarrow \frac{1}{2} \left|{\Dtarget}'-D\right|$\;
$\vec{v}_{ij} \leftarrow \left(\Qreduceddiff_{ij}\right)^2 \cdot \operatorname{sign}\left(\Qreduceddiff_{ij}\right) \cdot \vec{u}_{ij}$ for all $i\neq j$\;
\caption{Compute the applicable forces with an interaction-based cost. The $\operatorname{distances}$ function converts interactions $Q_{ij}$ into corresponding distances $Q_{ij}^{-1/6}$.}
\label{alg:applicable_forces}
\end{algorithm}

\subsection{Regularizing the temperatures}
\label{subsec:regularize-temperatures}

BLaDE stabilizes the iterations by rescaling the forces: for each pair $(i,j)$, the \textit{maximum temperature} is the largest safe multiplier $T_{ij} = |\Dwalk_{ij}| / \lVert\vec{v}_{ij}\rVert$, and we apply the \textit{global temperature} multiplier $T_{\mathrm{min}} = \min_{ij} T_{ij}$ on all forces to preserve their relative lengths and prevent each force from overshooting its target.

Optimizing interactions can make $T_{\mathrm{min}}$ a bottleneck: a few small maximum temperatures can turn it low and freeze other pairs with high temperature ratios $R_{ij}=T_{ij}/T_{\min}$, even with large corresponding $\Qdiff_{ij}$. Defining effective regulation cursors $c_{ij}\in[0,1]$, we compress the dynamic range of the ratios as $R^{\mathrm{reg}}_{ij}=R_{ij}^{\,1-c_{ij}}$, and we rescale the force vectors (leaving null vectors unchanged):
\begin{equation}
    \vec v^{\,\mathrm{reg}}_{ij}=R_{ij}/ R^{\mathrm{reg}}_{ij} \cdot \vec v_{ij}
    \label{eq:regulated_vectors}
\end{equation}
Since $1 \leq R^{\mathrm{reg}}_{ij} \leq R_{ij}$, each force can only remain constant or grow, without overshooting its target.

The cursor is set to $c_{ij}=g(\cglob,\cpair_{ij})\in[0,1]$ to combine a global regulation cursor $\cglob\in[0,1]$ (to choose how much regulation we apply) and a pairwise regulation cursor $\cpair_{ij} =\lVert{\vec{v}_{ij}}\rVert /\max_{i'j'} \lVert\vec{v}_{i'j'}\rVert \in[0,1]$ (to regulate important forces more). We set the modulation $g$ continuous, increasing in both arguments, and get the property that at fixed $\Dwalk_{ij}$, the order of the lengths of the forces is preserved.
We also set boundary conditions to give precedence to low regulation and for $\cglob$ to dominate each $\cpair = \cpair_{ij}$ on strong regulation:
\begin{equation}
\begin{aligned}
g(0,\cpair)&=0 &\qquad g(\cglob,0)&=0\\
g(1,\cpair)&=1 &\qquad g(\cglob,1)&=\cglob
\end{aligned}
\label{eq:g_boundary_conditions}
\end{equation}
To satisfy those, we define a smooth increasing map $\mu:[0,1]\to[0,1]$ and choose interior points of $g$ as:
\begin{equation}
    g(\cglob,\cpair)=\frac{\cglob\,\mu(\cpair)}{1+\cglob(\mu(\cpair)-1)}
    \label{eq:g_definition}
\end{equation}
\begin{equation}
    \mu(\cpair)=\frac{(\cpair)^s}{(\cpair)^s+\left(\frac{1}{\mumid}-1\right)(1-\cpair)^s}
    \label{eq:mu_definition}
\end{equation}
where $s>0$ controls the transition stiffness, and $\mu(0.5)=\mumid$ is set to $\frac{1}{1/\gmid -1}$ such that $g(0.5,0.5)=\gmid$.

\subsection{Optimizing the local equilibrium}
\label{subsec:optimize-local-equilibrium}

The FR method and its implementation spring layout do not ensure that each iteration decreases the cost and it can converge to a solution that is not even locally optimal.
Indeed, it applies pairwise attractive forces $f_a(D_{ij},w)=w_{ij}\cdot D_{ij}^2/k$ and repulsive forces $f_r(D_{ij})=-k^2/D_{ij}$ on pairs of points, where $w_{ij}$ and $k$ are constants that can be chosen. On each pair, the forces cancel at distance $k/\sqrt[3]{w}$, which enables to target $\Dtarget$ by choosing $k=1$ and $w_{ij}=1/(\Dtarget_{ij})^3$. However, the total force potential that is minimized is $\sum_{i<j}(-k^2\ln D_{ij}+w_{ij}D_{ij}^3/(3k))$ rather than $\|D^{\mathrm{diff}}\|_2$, because each force considers its cost decrease in isolation, disregarding the actual equilibrium between the forces when the points move more slowly.

To account for it and target a (possibly local) optimal, we define a reachable distance $\dmaxwalk$ and use Alg.~\ref{alg:target_local_equilibrium} to compute reachable target interactions $\Qtargetlimitedwalk$. Then, we use them as input in Alg.~\ref{alg:applicable_forces} so that only the expected improvements define the norms of the forces.

\begin{algorithm}
\KwIn{$\Qtarget$, $\Dtarget$, $D$, $\dmaxwalk$}
\KwOut{$\Qtargetlimitedwalk$}
\BlankLine
$\Ddiff \leftarrow \Dtarget - D$\;
$\Dreduceddiff \leftarrow \Ddiff$ clipped by $[-2\cdot \dmaxwalk\;;\; 2\cdot \dmaxwalk]$\;
$\Qtargetlimitedwalk \leftarrow \operatorname{weights}(D+\Dreduceddiff)$\;
\caption{Compute the new targets subject to limited distances to walk. The $\operatorname{weights}$ function converts distances $d$ into corresponding interactions $d^{-6}$.}
\label{alg:target_local_equilibrium}
\end{algorithm}

Clever values for $\dmaxwalk$ must also be chosen. The intuition is to set it to a high value for the first iterations to give the chance to all forces to reach their targets. It must then decrease to balance the magnitudes of the forces given the actual local improvements they can bring. It must also not converge too fast to 0 (e.g., exponentially) to prevent each iteration from having a corrective force negligible compared to the former one.

\subsection{Latent dimensions}
\label{subsec:latent-dimensions}

Force-directed layouts can get trapped in poor local optima in 2D. To escape these, BLaDE temporarily optimizes in higher-dimensional latent coordinates (transient auxiliary dimensions with no physical meaning), enabling exploration that does not rely on annealing effects in order to save runtime. Rather than computing a fixed high-dimensional embedding and projecting once, BLaDE can add or remove latent dimensions on the fly during iterations. When adding a dimension, new coordinates are initialized with small random values; under strong forces, these tend to amplify and help points move around other ones. When removing dimensions, BLaDE progressively flattens the space between parallel hyperplanes, clipping the selected coordinates over multiple iterations while continuing to apply forces. The space can first be rotated via Principal Component Analysis (PCA)\cite{scikit-learn} to flatten the directions that least affect the distances.

Figure \ref{fig:workflow} depicts the example of a dimension sequence $[4, 8, 2, 2]$, which defines three consecutive rounds, labeled by their dimension transitions $4\rightarrow8$ (adding dimensions), $8\rightarrow2$ (removing dimensions) and $2\rightarrow2$ (staying in 2 dimensions). Each round runs a fixed number of iterations. The sequence can be freely chosen, even using 1 dimension or finishing at a different number than 2. Intuitively, the more dimensions, the more likely points can escape from local optima. For instance, a point trapped in a triangle in 2D where it is being repulsed by the 3 points surrounding it can get around this obstruction in the 3rd dimension, and so can a point in a tetrahedron by going to the 4th dimension, etc. High-dimension iterations can however be slower due to more coordinates to compute, and it is necessary to eventually flatten coordinates back to the space of the desired dimensionality.

Adding dimensions to $n$ positions $\vec x_{i} \in \mathbb{R}^k$ is performed as follows: define the thickness $\Delta_r$ along each existing dimension $r$ as the third quartile minus the median of the coordinates, and the volume per point as $1/n\cdot\prod_r \Delta_r$. The corresponding hypercube edge length is the $k$-th root of the volume. Coordinates along the new dimensions are sampled from a uniform distribution centered at 0 and whose size is the edge length.

\subsection{Constraints satisfaction with scaling-in-the-loop}
\label{subsec:constraints-scaling}

BLaDE translates the distance-ratio constraint of Eq.~\ref{eq:max_min_distance_constraint} to add a degree of freedom $\lambda > 0$:
\begin{subequations}
\label{eq:distance_constraints}
\begin{align}
\|\vec{x}_i\| \le \dmax' = \kappa \cdot \lambda \cdot \dmax\quad \forall i \label{eq:distance_constraints_a}\\
D_{ij} \ge \dmin' = \lambda \cdot \dmin\quad\forall i\neq j \label{eq:distance_constraints_b}
\end{align}
\end{subequations}
where $\kappa > 0$ tunes the strictness of these constraints. Typically we set $\kappa > 1$ in early iterations for exploration, then $\kappa < 1$ to reduce stochastic violations. Violations of Eq.~\ref{eq:distance_constraints} are tackled through forces of Eq.~\ref{eq:force_constraints} to restore feasibility:
\begin{subequations}
\label{eq:force_constraints}
\begin{align}
\vec{f}^{\mathrm{\,max\,dist}}_{i} = & \min(0, \dmax' - \lVert\vec{x}_i\rVert) / \lVert\vec{x}_i\lVert \; \cdot \; \vec{x}_i \quad & \forall i \label{eq:force_constraints_a}\\
\vec{f}^{\mathrm{\,min\,dist}}_{ij} = & 1/2 \cdot \min(0, D_{ij} - \dmin') \; \cdot \; \vec{u}_{ij}\quad & \forall i \neq j \label{eq:force_constraints_b}
\end{align}
\end{subequations}
Note that those center the positions around $(0,0)$, and they should not be recentered using a barycenter as it generally differs from the center of the smallest enclosing circle.

A good $\lambda$ cannot be predetermined because it depends at each iteration on the positions of the points: when they are mostly limited by a too low $\dmax'$ (resp. a too high $\dmin'$), $\lambda$ should increase (resp. decrease).

To implement this behavior, $\lambda$ is initialized to make Eq.~\ref{eq:distance_constraints_b} a binding constraint. Then, at each iteration, the scaling factor $\dfactor$ on the positions that minimizes $\rVert \Qdiff \lVert$ of Eq.~\ref{eq:qdiff_ddiff} is computed, and $\lambda$ is updated through $\lambda := \dfactor \cdot \lambda$ to alleviate the (most) limiting bound of Eq.~\ref{eq:distance_constraints}. Defining the scaling factor $\qfactor=\dfactor^{-6}$ on interactions, the goal is to minimize:
\begin{equation}
\begin{aligned}
    & \|\Qtarget-\qfactor\cdot Q\|_2^2& \\
    & =  \qfactor^2\sum_{ij} Q_{ij}^2 - 2 \qfactor \sum_{ij} \Qtarget_{ij}Q_{ij} + \sum_{ij} \left(\Qtarget_{ij}\right)^2 &
\end{aligned}
\label{eq:scaling_cost}
\end{equation}
Minimizing this polynomial in $\qfactor$, we find:
\begin{equation}
    \dfactor = \sqrt[6]{\frac{\sum_{ij} Q_{ij}^2}{\sum_{ij} \Qtarget_{ij} Q_{ij}}}
    \label{eq:optimal_scaling}
\end{equation}
In practice, rare near-collisions can create large values in $\Qdiff$ that drive $\dfactor$ undesirably high and distort all positions. For robustness, define a nonnegative cap $ceiling$ as a chosen quantile of non-diagonal terms of $\Qdiff$ clipped positive. On pairs with $\Qdiff_{ij} > ceiling$, apply reduction weight $w_{ij}=ceiling/\Qdiff_{ij}$ (else $1$) and use the weighted scaling factor, to apply at each iteration on the positions and $\lambda$:
\begin{equation}
    \dfactor = \sqrt[6]{\frac{\sum_{ij} w_{ij} Q_{ij}^2}{\sum_{ij} w_{ij} \Qtarget_{ij} Q_{ij}}}
    \label{eq:optimal_weighted_scaling}
\end{equation}

\section{Numerical results}
\label{sec:numerical-results}

\begin{figure*}
    \centering
    \includegraphics[width=\linewidth]{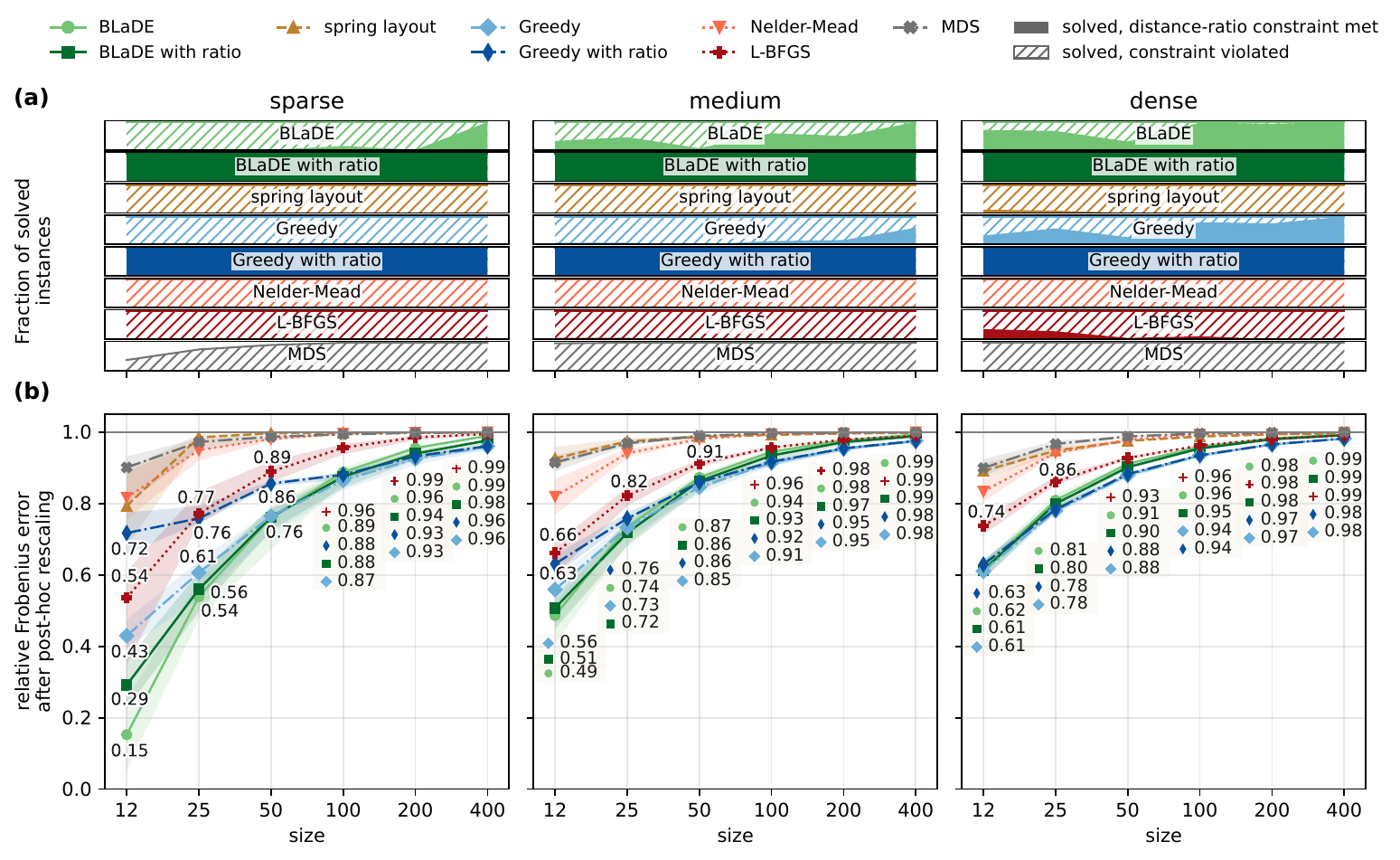}
    \caption{Embedding results for the weighted random dataset. The x-axes represent the sizes of the instances. (a) Fraction of solved instances. The only failures correspond to MDS on small disconnected graphs because of infinite distances. A plain color means it also follows the distance-ratio constraint. (b) Quality of the found solutions given by the relative Frobenius error $\frac{\lVert \Qtarget-Q\rVert_2}{\lVert \Qtarget\rVert_2}$, that we enhance with post-hoc rescaling. Curves correspond to medians, and shades materialize first and third quartiles. An error of $0$ corresponds to a perfect embedding, and $1$ can be naïvely obtained by making all positions far apart to let $Q$ have negligible values.
    }
    \label{fig:random_data_summary}
\end{figure*}

\subsection{Methods}
\label{subsec:methods}

We now present the numerical results from benchmarking BLaDE against various methods: spring layout, Greedy, Nelder-Mead, L-BFGS, MDS and GEAN. For each method, we use a large number of iterations to get best results under high budget. The other hyperparameters have been hand-selected based on a small subset of instances. It remains an open question to determine the optimal hyperparameters.

For NetworkX’s \texttt{spring\_layout} implementation (version 3.4.2) of the FR algorithm, we use 30,000 iterations and set \texttt{scale=None} to disable rescaling that is not adapted to the problem.

For SciPy's Nelder-Mead, we use 500,000 maximum evaluations (reached in practice for $n \geq 100$), and for L-BFGS, we use 100,000 maximum iterations and 1,000,000 maximum evaluations (never reached). We sample the initial coordinates uniformly in $[-1,1]$.

For Greedy, we use a triangular layout, create 6 times more sites than points to embed, and choose an inter-site spacing of $1$. We adapt it to minimize the same cost $\rVert \Qdiff \lVert_2$ as BLaDE. We also run a constrained ``Greedy with ratio'' version where only sites satisfying the distance-ratio constraint are used.

For scikit-learn's MDS (version 1.7.2), we use 100,000 maximum iterations. To prevent infinite distances that it does not support and mitigate triangular inequality violations to which it is sensitive, we preprocess the graphs as such: each edge is first weighted by its target distance, and then, weights are replaced by the shortest weighted paths.

GEAN, primarily designed for unweighted graphs, is reproduced with its paper's settings: a minimum pairwise distance of $4$, a maximum pairwise distance of $100$ (which leads to a maximum radial distance always lower than $50$ in practice after centering), hidden layers \verb|[64, 36, 18, 9, 18, 36, 64]| and providing input positions precomputed by 50 iterations of spring layout. The only change we make compared to the paper is that we return the found UD solution of the last iteration instead of performing an inference where the dropout layers could return invalid solutions. It increases the fraction of UD-valid solutions from 2.7\% to 34.7\% (the feasibility metric for the molecular instances in section~\ref{subsec:results}).

For BLaDE (corresponds to qoolqit version 1.2.0), we choose the transient dimensions as $[5, 4, 3, 2, 2, 2]$, $1000$ iterations per round (for a total of 5000) and enable PCA. Initial coordinates are uniformly sampled in $[-1,1]$. We fix $\qdifffactor = 0.9$, $\cglob = 0.1$, $\gmid=0.25$ and $s=1$. In addition to ``BLaDE'', we run a constrained variant ``BLaDE with ratio'' that enforces the distance-ratio constraint, by setting $\kappa = 2$ of section~\ref{subsec:constraints-scaling} in the first iteration,
decreasing it linearly to $0.94$ during the dimension transitions $5\rightarrow4\rightarrow3\rightarrow2$, and increasing it linearly back to $0.98$ during the remaining $2000$ iterations for the transitions $2\rightarrow2\rightarrow2$. We set a smoothly decreasing maximum distance to walk $\dmaxwalk(x) =2\dmax' \cdot(1 - \sin(x \cdot \pi / 2))$, where $x$ is the current iteration index divided by the total of $5000$, and $\dmax'$ is defined as in Eq.~\ref{eq:distance_constraints}a for ``BLaDE with ratio'' and $\dmax' = \max_{i\neq j} D_{ij}$ for ``BLaDE''. Intuitively, the first iterations can walk close to $2\dmax'$ and quickly scale positions up by orders of magnitude if needed. The sine function enables to start with a close-to-linear decay, as decaying faster would leave little budget for middle iterations, and decaying slower would force a sharper decrease later. The function  converges to 0 at $x=1$ with a null derivative: this ensures that the final equilibrium corresponds to a null maximum distance to walk, and that it converges slowly enough to prevent the situation where each iteration goes beyond what the subsequent iterations can correct.
We set the cap ceiling of section~\ref{subsec:constraints-scaling} to the percentile $100-20/(n-1)$ (with $n$ the size of the instance), to have a percentile index sufficiently lower than $100$, but still increasing with $n$ as random high interactions become a smaller fraction, and clip the scaling factor between $0.1$ and $10$ to limit too quick variations.

While ``BLaDE with ratio'' and ``Greedy with ratio'' naturally center positions around $(0,0)$, other methods do not, so we check the distance ratio of each method by centering the solution on the center of the smallest enclosing circle.

\begin{figure*}
    \centering
    \includegraphics[width=\linewidth]{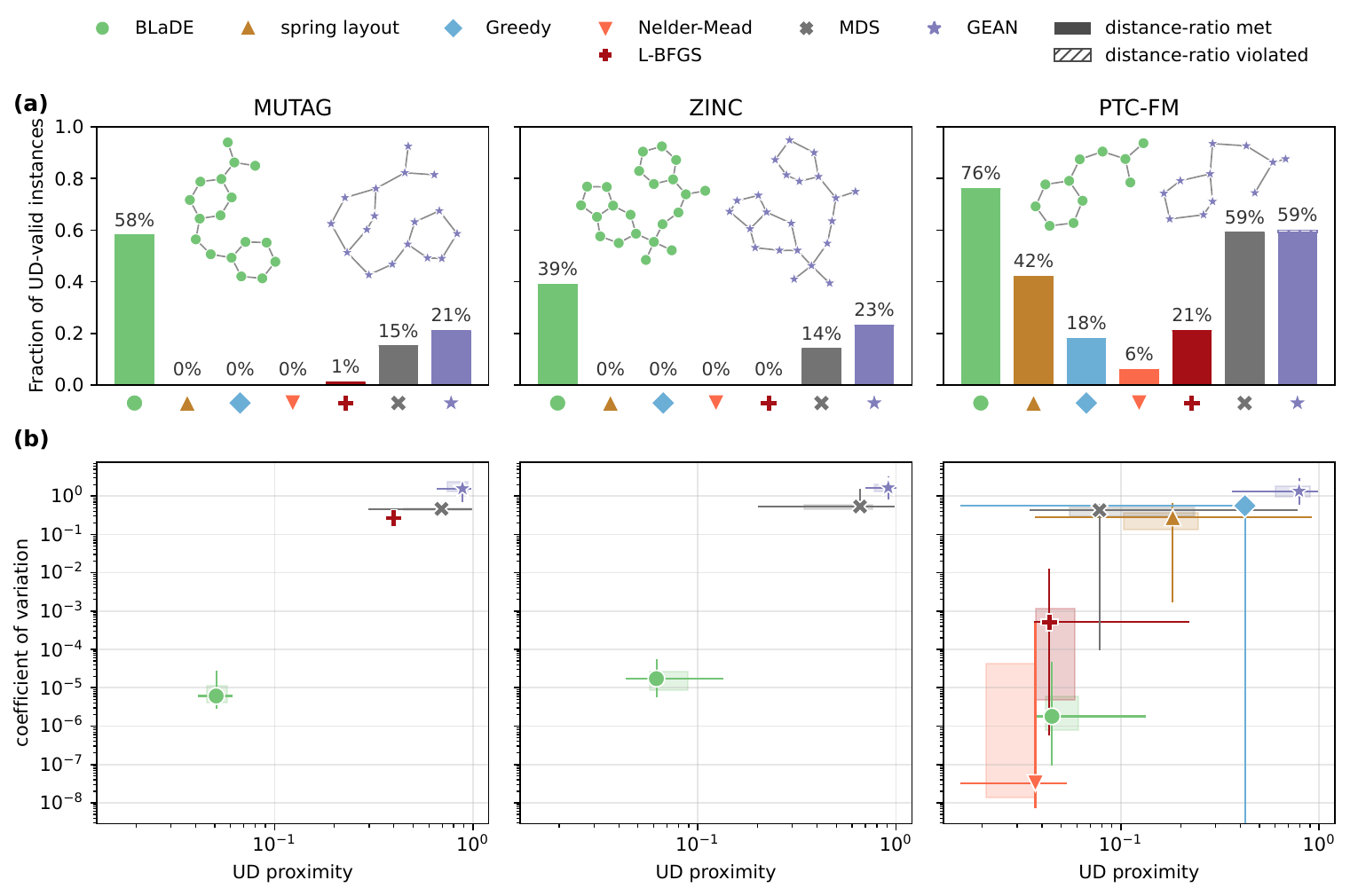}
    \caption{Embedding results for the molecular datasets. (a) Fractions of UD-valid instances. On the considered instances, the distance ratio has additionally always been met for free, except on one instance for GEAN due to a too short pair. As an illustration, examples of UD-valid solutions are also drawn in each dataset for the 2 best methods. (b) Quality of the UD solutions from Eq.~\ref{eq:ud_quality} as 2-dimensional log-log boxplots. The lower, the better. Only UD-valid solutions are included, and methods that did not return any are omitted.
    }
    \label{fig:molecular_summary}
\end{figure*}

\subsection{Benchmarks}
\label{subsec:results}

We generate symmetric target interaction matrices $\Qtarget\in\mathbb{R}^{n\times n}$ with $\Qtarget_{ii}=0$, considering sizes $n\in\{12,25,50,100,200,400\}$ and three density classes:
(i) sparse $\rho\in[0.1,0.3)$,
(ii) medium $\rho\in[0.3,0.7)$,
(iii) dense $\rho\in[0.7,1.0]$,
where the density is the fraction of nonzero off-diagonal values $\Qtarget_{ij}\ge 0$ uniformly sampled from $[0, 1]$ for all $i\neq j$.
For each graph size and  density, we generate 50 instances.
Finally, to benchmark the methods under a demanding distance-ratio constraint, we set the constraint of each instance from the most compact arrangement of the same number of points: the hexagonal-shaped triangular lattice of Pulser's \verb|TriangularLatticeLayout|~\cite{pasqal_pulser}. Its distance ratio, multiplied by $1.2$ to leave a small margin, is used as the distance-ratio constraint.

For weighted random instances, we perform a post-hoc rescaling on the output of each method using Eq.~\ref{eq:optimal_scaling}. Without it, spring layout had a median relative Frobenius error of $728$, and MDS of $30300$ (aggregated on all sizes and densities), far above $1$. The rescaling brings them below $1$. It also improves the results of ``Greedy with ratio'', going from a median of $0.932$ to $0.882$. For the other methods, the gain is marginal, below $0.006$.

Figure \ref{fig:random_data_summary} reports the corresponding results. The subplot \ref{fig:random_data_summary}a shows the fraction of solved instances, always at 100\% except for MDS on small disconnected graphs. The plain colored regions indicate embeddings that also satisfy the imposed distance-ratio constraint. BLaDE and Greedy tend to naturally satisfy the distance-ratio constraint on denser and larger instances, especially BLaDE, and the constrained versions always satisfy it. The other methods rarely meet the distance-ratio constraint, as they do not include the constraint, and do not get it for free either.

The subplot \ref{fig:random_data_summary}b compares the relative Frobenius error between $\Qtarget$ and $Q$ for each instance, where lower values indicate better embeddings. On the low-error solutions, which are the most likely to have a useful implementation in the Hamiltonian, BLaDE precisely provides the best accuracies, i.e., whenever the best methods achieve errors below $0.6$, Greedy arriving second. It does so by far: on the smallest sparse instances, it finds $0.15$ against $0.43$ for Greedy ($-64\%$ of error), and $0.29$ against $0.72$ with the ratio constraint ($-59\%$ of error).
The constrained version of Greedy specifically struggles to be on-par with BLaDE, and on sparse instances, it can only achieve so for sizes $n \geq 100$. On denser and larger instances, BLaDE and Greedy are almost on-par, with a tiny advantage for Greedy of up to $0.03$ ($-3.7\%$ of error compared to BLaDE). MDS, spring layout and the black-box methods do not compete well. The comparison between BLaDE and spring layout whose error starts at $0.79$ validates the high improvement brought by BLaDE's specific features.

\begin{table}[t]
  \centering
  \caption{Molecular datasets: interquartile ranges of the instance sizes $n$ and of the edge densities
           $\rho = |\mathcal{E}| / \binom{n}{2}$, over the 100 molecules drawn from each subset.}
  \label{tab:molecular}
  \begin{tabular}{lrr}
    Subset & Size (Q1--Q3) & Density $\rho$ (Q1--Q3) \\
    \midrule
    MUTAG  & 14--22 & 0.11--0.16 \\
    ZINC   & 19--25 & 0.09--0.11 \\
    PTC-FM & 7--19  & 0.12--0.29 \\
  \end{tabular}
\end{table}

We then evaluate the methods on unweighted molecular graph datasets: MUTAG, ZINC and PTC-FM. These datasets are standard benchmarks in graph machine learning and are distributed through the PyTorch Geometric package~\cite{fey2019pyg}. For each dataset, we randomly sample 100 molecular graphs and extract their connectivity structure. Each molecule is represented as an unweighted graph by setting $Q_{ij}=1$ for edges $(i,j)\in \mathcal{E}$ and $Q_{ij}=0$ for non-edges $(i,j)\notin \mathcal{E}$. The features of the resulting graphs are summarized in Table \ref{tab:molecular}. In this benchmark, BLaDE is run without explicitly enforcing the distance-ratio constraint, but we still check whether its solutions satisfy the ratio $50/4=12.5$ used in the GEAN setting. We report two UD quality indicators:

\begin{equation}
\begin{aligned}
\text{UD proximity} &= \frac{\max Q_{(i,j)\notin\mathcal{E}}}{\min Q_{(i,j)\in\mathcal{E}}} \\
\text{coefficient of variation} &= \frac{\operatorname{std}Q_{(i,j)\in\mathcal{E}}}{\operatorname{mean} Q_{(i,j)\in\mathcal{E}}}
\end{aligned}
\label{eq:ud_quality}
\end{equation}
The first quantity compares the strongest non-edge interaction to the weakest edge one (UD-valid if below 1).
The second quantity measures how uniform the intended edge interactions are.

The Figure \ref{fig:molecular_summary}a reports the fraction of UD-valid solutions and shows that BLaDE has the highest success on all datasets. On MUTAG and ZINC, it produces $1.7$ to $2.8$ times as many UD-valid solutions as the best other method, which is our improved GEAN. This is a notable result because BLaDE is not explicitly designed as a UD graph embedder: it minimizes $\lVert \Qdiff \rVert$, yet this cost naturally encourages connected pairs to be close and non-connected pairs to be sufficiently separated. The plots also show that the UD-valid solutions almost always satisfy the GEAN distance-ratio constraint as well.

The Figure \ref{fig:molecular_summary}b analyzes the quality of the UD-valid solutions using the two indicators above. The two-dimensional log–log boxplots show that BLaDE’s UD-valid solutions are not merely feasible; they are also of substantially higher quality. In MUTAG and ZINC, BLaDE achieves both lower UD proximity (tenfold) and lower coefficient of variation (by 5 orders of magnitude) than other methods, meaning that non-edge interactions remain well separated from edge interactions while the edge interactions themselves are relatively uniform. This gives a stronger and more stable physical encoding of the graph structure on the neutral atom quantum computer. In PTC-FM, Nelder-Mead achieves competitive qualities, but only on its much lower success rate: $6\%$ versus $76\%$ for BLaDE. BLaDE therefore provides a better trade-off between reliability and quality, producing many more valid embeddings while maintaining strong separation and uniformity.

Overall, these numerical experiments show that BLaDE’s design choices translate into practical advantages on both weighted and unweighted targets. On weighted random matrices, the method is distinguished by its ability to jointly satisfy hardware-inspired distance constraints and minimize the interaction-level error. On molecular graphs, it transfers effectively to the UD setting and produces both valid and high quality embeddings more often than other baselines.

\section{Conclusion}
\label{sec:conclusion}

We introduced BLaDE, a force-directed embedding algorithm designed to map target interaction matrices onto neutral-atom quantum processors. Unlike distance-based layout methods, BLaDE directly optimizes the interactions, which are the quantities that determine the implemented Rydberg Hamiltonian. By combining balanced force updates, local-equilibrium optimization, temporary latent dimensions, and a scaling-in-the-loop mechanism for hardware constraints, the method addresses several practical difficulties of register design: non-convexity, sensitivity to local minima, infeasible geometries, and degradation under realistic distance-ratio constraints.

Across weighted random instances and molecular graph benchmarks, we enhance the baselines, and BLaDE outperforms them by greatly improving the quality of the better solutions, and staying on-par for the worse solutions from denser and larger instances, while reliably satisfying the imposed distance-ratio constraint. On unweighted molecular graphs, BLaDE produces substantially more unit-disk embeddings and in most cases improves their quality by orders of magnitude. Finally, our new method provides a flexible framework that can output 3D positions, and allows extensions to alternative interaction laws such as XY-based Hamiltonian, in which $Q_{ij}$ scales as $D_{ij}^{-3}$ .

\section*{Author statement}

Clément de Terrasson de Montleau designed and developed the method, carried out the experiments, and wrote the manuscript. Victor Drouin-Touchette contributed to the method design and helped with the final manuscript writing. Wesley Coelho contributed to the method design, helped with the final manuscript writing and supervised the project.

\bibliographystyle{IEEEtran}
\bibliography{references}

\end{document}